\documentstyle[11pt]{article}

\newtheorem{thm}{Theorem}[section]

\newcommand{\A}{\mathcal{A}}

\begin{document}   
\title{\bf  {Gravitational Energy in  Van Stockum  Space-Time}}
\author{{ Ragab M. Gad$^{1,2}$ \thanks{%
E-mail: ragab2gad@hotmail.com} \,\, and H. A. Alharbi$^{1}$ \thanks{%
E-mail: haifaa.alharbi135@gmail.com }}\\
\newline
 {\it $^1$
 Department of Mathematics, Faculty of Science, University of Jeddah,
}\\
 {\it 21589 Jeddah, Saudi Arabia}
 \\
{\it $^2$ Department of Mathematics, Faculty of Science, Minia University,}\\
 {\it   61915 El-Minia,  Egypt}
}

\date{\small{}}

\maketitle
\begin{abstract}
The purpose of this paper is to illustrate the problem of energy and momentum distributions of Van Stockum space-time within the framework of two different theories of gravity, general relativity and teleparallel gravity.
 We have shown that for all homogeneous space-times with metric components $g_{\mu\nu}$ being functions of time variable, $t$, alone and independent of space variables the total gravitational energy for any finite volume is identically zero.
By working with general relativity, we have calculated the energy-momentum density  for Van Stockum space-time  using  double index complexes and in the framework  teleparallel gravity, we used the energy-momentum complexes of Einstein, Bergmann-Thomson and Landau-Lifshitz. In our analysis, we sustained  that general relativity and teleparallel gravity are equivalent theories of space-time under consideration. For space-time under consideration, we have shown that different complexes of  energy-momentum density do not provide the same results neither in  general relativity nor in teleparallel gravity.

\end{abstract}

\setcounter{equation}{0}
\section{Introduction}

Although there are some basic conceptual differences between the theory of general relativity  and the theory of teleparallel gravity, they are theoretically equivalent theories. These differences may appear in the fact that in the case  of general relativity, gravity curves the space-time and forms the geometry, that is, the curvature is the geometry of gravity. In the theory of teleparallel gravity there are no geodesic equations, because the gravity is described by torsion, not by geometrization as in  general relativity, as a force acting on a test particle \cite{HS79}.  The most important difference is that the tetrad field in TPG, instead of the metric tensor, $g_{\mu\nu}$, in GR, is the primary geometric object.
\par
Since the beginning of general relativity, the subject of gravitational energy and momentum distributions is one of the  interesting and challenging problems. In attempts to get an acceptable  definition of this problem, various definitions have been provided \cite{E15}-\cite{W72}.
These prescriptions are limited to calculating the energy and momentum distributions in quasi-Cartesian coordinates, except for the prescription of M{\o}ller \cite{E1}, in order to obtain a  meaningful result. Despite these doubts, the definitions given have yielded acceptable results. For a certain space-time, some obtained results lead to the conclusion that these complexes  give the same
energy-momentum density \cite{Esame}-\cite{R06}. However, some examples of space-times  do not agree with these results \cite{Enot}-\cite{B92}.
\par
The theory of teleparallel gravity reproduces the empirical content of general relativity, but in a more similar format to the gauge theories of the standard model than gravity does. Therefore, this theory allows the treatment of energy-momentum localization to be addressed more satisfactorily \cite{M62}.
Many authors have used this theory to address many of the problems that have appeared in the  theory of general relativity, which considers energy-momentum distribution  to be one of these problems.
\par
In the context of this theory, there have been many attempts to get a definition
of the issue of energy and momentum distributions for a  space-time. The first attempt, after Einstein suggested this theory, was by
M{\o}ller \cite{M62}, who observed that the tetrad's description of the gravitational field addresses the problem of localizing energy and momentum.
The topic of energy and momentum distributions  has been extensively studied in the theory of teleparallel  gravity and there are many accepted results on this topic \cite{Gad08}-\cite{CW95}.
\par
This paper is organized as follows: in section 2,  we use the double index complexes to calculate the energy-momentum density of  Einstein, M{\o}ller, Bergmann-Thomson and Landau-Lifshitz.
 In section 3 we summarize the general morphology of the basic
concepts  of teleparallel gravity.  Section 4, deals with the calculation of the energy-momentum
density for the given space-time. In section 5,  we give discussion and
conclusions.
\setcounter{equation}{0}
\section{ Energy-momentum complexes in General Relativity}
In the theory of General Relativity, the energy-momentum prescriptions  of Einstein (E), M{\o}ller (M), Bergmann-Thomson (BT) and Landau-Lifshitz (LL)  can be written as follows:
\begin{equation} \label{1}
\gamma^k_i=\partial_\ell U^{[k\ell]}_i,  \qquad \gamma^{ik}=\partial_\ell U^{i[k\ell]},
\end{equation}
with
\begin{equation} \label{2}
U^{[k\ell]}_i=\left\lbrace
\begin{array}{c l}
\frac{1}{2\kappa}(-g)^{-\frac{1}{2}}g_{in}\partial_{m}[(-g)(g^{kn}g^{\ell m}-g^{\ell n}g^{km})] & (E),\\
\frac{1}{\kappa}(-g)^{\frac{1}{2}}g^{km}g^{\ell n}(\partial_m g_{in}-\partial_n g_{im}) & (M).
\end{array}
\right.
\end{equation}

\begin{equation} \label{3}
U^{i[k\ell]}=\left\lbrace
\begin{array}{c l}
\frac{1}{2\kappa}(-g)^{-\frac{1}{2}} \partial_{m}[(-g)(g^{ik}g^{\ell m}-g^{i\ell}g^{km})] & (BT),\\
\frac{1}{2\kappa}\partial_{m}[(-g)(g^{ik}g^{\ell m}-g^{i\ell}g^{km})] & (LL),
\end{array}
\right.
\end{equation}
where $\kappa=8\pi$.\\
These complexes are called {\bf{double index complexes}} \cite{C66}.\\
The superpotential $U^{[k\ell]}_i$ or $U^{i[k\ell]}$ is made up from the contravariant components of the metric tensor and their derivative. By definition, each of them is anti-symmetric in the pair of indices $k$ and $\ell$, hence the complexes satisfy the local conservation law
$$
\partial_k\gamma^k_i=0, \qquad \partial_k\gamma^{ik}=0.
$$
$\gamma^0_0$, $\gamma^{00}$, and $\gamma^0_\mu$, $\gamma^{\mu 0}$ are, respectively,  the energy and momentum density components.The complexes deriving from superpotential (\ref{3}), are symmetrical in the two indices $i$ and $k$; this permits the establishment of a conservation law of angular momentum as well.
\par
The energy and momentum distributions in the above various complexes are defined by
\begin{equation} \label{4}
P_i=\int\int\int \gamma^0_i d^3x=\int_{S}\int U^{[0\ell]}_i dS_\ell,
\end{equation}
\begin{equation} \label{5}
P^i=\int\int\int \gamma^{i0} d^3x=\int_{S}\int U^{i[0\ell]} dS_\ell,
\end{equation}
where the first integration in both two above equations is carried out on the surface $x^0= const.$ and the second on a close two-surface $S$ belonging to the same hypersurface and expanding to infinity.
\par
Now, we assume the general metric
\begin{equation} \label{6}
ds^2=g_{\mu\nu} dx^\mu dx^\nu,
\end{equation}
where the metric components $g_{\mu\nu}$ are functions of time variable, $t$, alone and independent of space variables. Consequently, the superpotential $U^{[k\ell]}_i$ and $U^{i[k\ell]}$ are also functions of the time variable alone. From (\ref{1}) and the antisymmetric of superpotential, we have
\begin{equation} \label{7}
\gamma^0_0=0,  \quad \gamma^{00}=0.
\end{equation}
Therefore, the following is then obvious
\begin{thm} \label{th}
For all homogeneous space-times anzatz with the metric (\ref{6}) the total gravitational energy for any finite volume is identically zero.
\end{thm}

One of the exact solutions to Einstein's equations that were found somewhat early is the Van Stockum space-time. In this space-time, the gravitational field is generated by the dust that is rotating around the axis of cylindrical symmetry. \\
It is presented by the following model, in Cartesian coordinates, \cite{Van37}
\begin{equation}\label{VS}
ds^2= fdt^2 -e^{2\psi}(dx^2 +dy^2)- \ell dz^2 + 2mdzdt,
\end{equation}
where $f, \ell, m$ and $\psi$ are functions of the variables $x$ and $t$.\\

\subsection{Einstein prescription}
The non-vanishing superpotential components, $U^{[k\ell]}_i$, of the Einstein complex of the line element (\ref{VS}), using equation (\ref{2}), are
\begin{equation}\label{Einstein}
\begin{array}{ccc}
^EU^{[01]}_0& =& -\frac{1}{2\kappa\sqrt{f\ell+m^2}}\big(2\psi^\prime(f\ell+m^2)+f\ell^\prime+mm^\prime\big),\\
^EU^{[03]}_0& =& \frac{1}{\kappa\sqrt{f\ell+m^2}}\big( 2m\dot{\psi}e^{2\psi}\big), \\
^EU^{[01]}_1& =& -\frac{1}{2\kappa\sqrt{f\ell+m^2}}\big(2\dot{\psi}\ell+\dot{\ell}\big)e^{2\psi},\\
^EU^{[02]}_2& =&-\frac{1}{2\kappa\sqrt{f\ell+m^2}}\big(2\dot{\psi}\ell+\dot{\ell}\big)e^{2\psi},\\
^EU^{[01]}_3& =& \frac{1}{2\kappa\sqrt{f\ell+m^2}}\big(\ell m^\prime - m\ell^\prime\big),\\
^EU^{[03]}_3& =& -\frac{1}{2\kappa\sqrt{f\ell+m^2}}\ell\dot{\psi}e^{2\psi}.
\end{array}
\end{equation}

Inserting the above components into (\ref{1}), we find the energy-momentum density in the sense of Einstein, respectively, as follows

\begin{equation}\label{Einstein-EM}
\begin{array}{ccc}
^E\gamma^0_0 =-\frac{1}{4\kappa ({f\ell+m^2})^{\frac{3}{2}}}\Big((f\ell+m^2)(4(f\ell+m^2)\psi^{\prime\prime}+2mm^{\prime\prime}+4mm^\prime\psi^\prime+f^\prime\ell^\prime+2f\ell^{\prime\prime}\\
+2\ell f^\prime\psi^\prime+2f\ell^\prime\psi^\prime)-f^2\ell^{\prime 2}+m^2 f^\prime\ell^\prime+2f\ell m^{\prime 2}-3fmm^\prime\ell^\prime-\ell f^\prime mm^\prime\Big),\\
^E\gamma^0_1 =-\frac{e^{2\psi}}{4\kappa ({f\ell+m^2})^{\frac{3}{2}}}\Big(2(f\ell+m^2)(4\ell\dot{\psi}\psi^\prime+\ell^\prime\dot{\psi}+2\dot{\ell}\psi^\prime+2\ell\dot{\psi}^\prime+\dot{\ell}^\prime)+2m^2\ell^\prime\dot{\psi}-\\
 2\ell^2\dot{\psi}f^\prime-4\ell mm^\prime\dot{\psi}-f\dot{\ell}\ell^\prime-\ell\dot{\ell}f^\prime-2\dot{\ell}mm^\prime\Big),\\

^E\gamma^0_3 =\frac{1}{4\kappa ({f\ell+m^2})^{\frac{3}{2}}}\Big(2(f\ell+m^2)(\ell m^{\prime\prime}-m\ell^{\prime\prime})+fm\ell^{\prime 2}+m\ell\ell^\prime f^\prime+2m^2 m^\prime\ell^\prime\\
-f\ell\ell^\prime m^\prime -\ell^2 f^\prime m^\prime -2\ell mm^{\prime 2}\Big).
\end{array}
\end{equation}

\subsection{M{\o}ller prescription}

The non-vanishing superpotential components, $U^{[k\ell]}_i$, of the M{\o}ller definition of the line element (\ref{VS}), using equation (\ref{2}), are
\begin{equation}\label{Moller}
\begin{array}{ccc}
^MU^{[01]}_0& =& \frac{\ell f^\prime}{\kappa\sqrt{f\ell+m^2}},\\

^MU^{[01]}_1& =& \frac{2\ell \dot{\psi}}{\kappa\sqrt{f\ell+m^2}}e^{2\psi},\\
^MU^{[02]}_2& =&\frac{2\ell \dot{\psi}}{\kappa\sqrt{f\ell+m^2}}e^{2\psi},\\
^MU^{[01]}_3& =& \frac{\ell\dot{m}}{\kappa\sqrt{f\ell+m^2}},\\
^MU^{[03]}_3& =& -\frac{f\ell\dot{\ell}}{\kappa(f\ell+m^2)^{\frac{3}{2}}}e^{2\psi}.
\end{array}
\end{equation}

Inserting the above components into (\ref{1}), we get the M{\o}ller's energy-momentum density, respectively, as follows

\begin{equation}\label{Moller-EM}
\begin{array}{ccc}
^M\gamma^0_0 =\frac{1}{2\kappa ({f\ell+m^2})^{\frac{3}{2}}}\Big((f\ell+m^2)(2\ell f^{\prime\prime}+f^\prime\ell^\prime)+m^2f^\prime\ell^\prime-\ell^2 f^{\prime 2}-2\ell mf^\prime m^\prime\Big),\\
^M\gamma^0_1 =\frac{e^{2\psi}}{2\kappa ({f\ell+m^2})^{\frac{3}{2}}}\Big((f\ell+m^2)(4\ell\dot{\psi}\psi^\prime +\dot{\psi}\ell^\prime+2\ell\dot{\psi}^\prime)-\ell^2\dot{\psi} f^\prime +m^2\dot{\psi}\ell^\prime-2\ell mm^\prime\dot{\psi}\Big),\\

^M\gamma^0_3 =\frac{1}{2\kappa ({f\ell+m^2})^{\frac{3}{2}}}\Big((f\ell+m^2)(2\ell\dot{m}^\prime+\dot{m}\ell^\prime)-\ell^2\dot{m} f^\prime-2\ell m\dot{m}m^\prime+m^2\dot{m}\ell^\prime\Big).
\end{array}
\end{equation}

\subsection{Bergmann-Thomson prescription}

The non-vanishing superpotential components, $U^{i[k\ell]}$, of the Bergmann-Thomson complex of the line element (\ref{VS}), using equation (\ref{3}), are
\begin{equation}\label{BT}
\begin{array}{ccc}
^{BT}U^{[001]}& =& -\frac{2\ell\psi^\prime+\ell^\prime}{2\kappa\sqrt{f\ell+m^2}},\\
^{BT}U^{[101]}& =& -\frac{2\ell\dot{\psi}+\dot{\ell}}{2\kappa\sqrt{f\ell+m^2}},\\
^{BT}U^{[202]}& =& -\frac{2\ell\dot{\psi}+\dot{\ell}}{2\kappa\sqrt{f\ell+m^2}},\\
^{BT}U^{[301]}& =& -\frac{2m\psi^\prime+m^\prime}{2\kappa\sqrt{f\ell+m^2}},\\
^{Bt}U^{[303]}& =& -\frac{e^{2\psi}}{\kappa(f\ell+m^2)^{\frac{5}{2}}}(-2m^4\dot{\psi}+\ell m^2\dot{f}+m^2 f\dot{\ell}-2f\ell m\dot{m}+2f^2\ell^2\dot{\psi}).
\end{array}
\end{equation}

Inserting the above components into (\ref{1}), we get the  energy-momentum density using the Bergmann-Thomson's prescription, respectively, as follows

\begin{equation}\label{BT-EM}
\begin{array}{ccc}
^{BT}\gamma^{00} =\frac{1}{4\kappa ({f\ell+m^2})^{\frac{3}{2}}}\Big(-2(f\ell+m^2)(\ell^\prime\psi^\prime +2\ell\psi^{\prime\prime}+\ell^{\prime\prime})-2m^2\ell^\prime\psi^\prime+2\ell^2 f^\prime\psi^\prime+4\ell mm^\prime\psi^\prime\\
 +f\ell^{\prime 2}+f^\prime\ell\ell^\prime+2\ell^\prime mm^\prime\Big),\\
^{BT}\gamma^{10} =\frac{1}{4\kappa ({f\ell+m^2})^{\frac{3}{2}}}\Big(2(f\ell+m^2)(\ell^\prime\dot{\psi} +2\ell\dot{\psi}^\prime+\dot{\ell}^\prime)+2m^2\ell^\prime\dot{\psi}-2\ell^2 f^\prime\dot{\psi}-4\ell mm^\prime\dot{\psi}\\
-f\ell^\prime\dot{\ell}-f^\prime\ell\dot{\ell}-2mm^\prime\dot{\ell}\Big),\\

^{BT}\gamma^{30} =\frac{1}{4\kappa ({f\ell+m^2})^{\frac{3}{2}}}\Big(-2(f\ell+m^2)(2m\psi^{\prime\prime}+m^{\prime\prime}) +2fm\ell^\prime\psi^\prime+2m\ell f^\prime\psi^\prime-4f\ell m^\prime\psi^\prime\\
+f\ell^\prime m^\prime+\ell f^\prime m^\prime+2mm^{\prime 2}\Big).
\end{array}
\end{equation}

\subsection{Landau-Lifshitz complex}
The non-vanishing superpotential components, $U^{i[k\ell]}$, of the Landau-Lifshitz prescription of the line element (\ref{VS}), using equation (\ref{3}), are

\begin{equation}\label{LL}
\begin{array}{ccc}
^{LL}U^{[001]}& =& -\frac{2\ell\psi^\prime+\ell^\prime}{2\kappa}e^{2\psi},\\
^{LL}U^{[101]}& =& \frac{2\ell\dot{\psi}+\dot{\ell}}{2\kappa}e^{2\psi},\\
^{LL}U^{[202]}& =& \frac{2\ell\dot{\psi}+\dot{\ell}}{2\kappa}e^{2\psi},\\
^{LL}U^{[301]}& =& -\frac{2m\psi^\prime+m^\prime}{2\kappa}e^{2\psi},\\
^{LL}U^{[303]}& =& -\frac{e^{4\psi}}{\kappa(f\ell+m^2)^{2}}(-2m^4\dot{\psi}+\ell m^2\dot{f}+m^2 f\dot{\ell}-2f\ell m\dot{m}+2f^2\ell^2\dot{\psi}).
\end{array}
\end{equation}

Using the above components in (\ref{1}), we obtain the Landau-Lifshitz energy-momentum density, respectively, as follows

\begin{equation}\label{LL-EM}
\begin{array}{ccc}
^{LL}\gamma^{00} =-\frac{e^{2\psi}}{2\kappa}\Big(4\ell\psi^{\prime 2}+2\ell\psi^{\prime\prime}+4\ell^\prime\psi^\prime+\ell^{\prime\prime}\Big),\\
^{LL}\gamma^{10} =\frac{e^{2\psi}}{2\kappa}\Big(4\ell\dot{\psi}\psi^\prime+2\ell\dot{\psi}^\prime+ 2\ell^\prime\dot{\psi}+2\dot{\ell}\psi^\prime+\dot{\ell}^\prime\Big),\\
^{LL}\gamma^{30} =-\frac{e^{2\psi}}{2\kappa}\Big(4m\psi^{\prime 2} +2m\psi^{\prime\prime}+4m^\prime\psi^\prime+m^{\prime\prime}\Big).
\end{array}
\end{equation}

\setcounter{equation}{0}
\section{ Teleparallel Gravity}
The name of teleparallel gravity is usually used to denote the
general three-parameter theory \cite{HS79}.
The theory of teleparallel gravity  equivalent to the theory of general relativity \cite{H02} can
actually be understood as a gauge theory for the translation group
based on Weitzenb\"{o}ck geometry \cite{Wei}.\\

\subsection{Fundamental concepts}
Let's start by reviewing the basic concepts of the theory of teleparallel gravity (see for example \cite{HS79,AP95,AP98,AGP00}.\\
 In the framework of teleparallel gravity the tetrad field, $h^{a}_{\,\,\,\mu}$, is used instead of the metric tensor, $g_{\mu\nu}$, in general relativity. A certain metric tensor admits many tetrad fields.
 There are many ways to create a tetrad, $h^{a}_{\,\,\,\mu}$, that corresponds to the metric tensor, $g_{\mu\nu}$.
  For example, it can be expressed by the dual basis of the differential one-form by choosing a coframe of the coordinate system \cite{HN90}. Another way to build a tetrad field is by finding a solution to the next relationship
\begin{equation}\label{g}
g_{\mu\nu}=\eta_{ab}h^a_{\,\,\mu}h^b_{\,\,\nu},
\end{equation}
where $\eta_{ab}= diag (1, -1, -1, -1)$.\\
 $h^a_{\,\,\mu}$ and its inverse, $h_a^{\,\,\nu}$,   should satisfy the  relationships
\begin{equation}\label{h}
h^a_{\,\,\mu}h_a^{\,\,\nu} = \delta_{\mu}^{\nu}; \quad  h^a_{\,\,\mu}h_b^{\,\,\mu} = \delta_{b}^{a},
\end{equation}
and
\begin{equation}\label{h-g}
h=\det (h^a_{\,\,\mu})=\sqrt{-g},
\end{equation}
where $g=\det( g_{\mu\nu})$ and $h=\det(h^{a}_{\,\,\,\mu})$.\\
Using the Minkowski metric, $\eta_{ab}$, the tangent space indices are raised
and lowered while by using the space-time metric, $g_{\mu\nu}$,
the space-time indices are raised and lowered.
 The parallel transport of  $h^{a}_{\,\, \nu}$ between two
neighboring points is coded into the covariant derivative
$$
\nabla_{\mu} h^{a}_{\,\, \nu} =\partial_{\mu} h^{a}_{\,\, \nu} -
\Gamma^{\alpha}_{\mu\nu}h^{a}_{\,\, \alpha},
$$
where
\begin{equation}\label{conn}
\Gamma^{\alpha}_{\,\,\mu\nu}= h_{a}^{\,\, \alpha}\partial_{\nu}
h^{a}_{\,\, \mu}=-h^{a}_{\,\, \mu}\partial_{\nu}h_{a}^{\,\, \alpha}
\end{equation}
is a Weitzenb\"{o}ck connection \cite{Wei}.  This connection provides a torsion that is defined as follows
\begin{equation}\label{tor}
T^{\rho}_{\,\,\mu\nu}=\Gamma^{\rho}_{\,\,\nu\mu} -
\Gamma^{\rho}_{\,\,\mu\nu}=h_{a}^{\,\, \alpha}(\partial_{\mu}
h^{a}_{\,\, \nu}-\partial_{\nu} h^{a}_{\,\, \mu} ),
\end{equation}
it is not symmetric in  $\mu$ and $\nu$.\\
The  connection $\Gamma^{\rho}_{\,\, \mu\nu}$ and the Levi-Civita
connection $\tilde{\Gamma}^{\rho}_{\,\, \mu\nu}$\footnote{
$\tilde{\Gamma}^{\rho}_{\,\, \mu\nu}
=\frac{1}{2}g^{\rho\sigma}(g_{\mu\sigma,\nu}+g_{\nu\sigma,\mu}
-g_{\mu\nu,\sigma})$} are connected by the following relation
$$
\Gamma^{\rho}_{\,\, \mu\nu}=\tilde{\Gamma}^{\rho}_{\,\, \mu\nu}+
K^{\rho}_{\,\, \mu\nu},
$$
where
$$
K^{\rho}_{\,\, \mu\nu}= \frac{1}{2}\big(T^{\,\,\rho}_{\mu\,\,
\nu}+T^{\,\,\rho}_{\nu\,\, \mu}- T^{\rho}_{\,\, \mu\nu}\big)
$$
is called the {\it{contortion tensor}}.\\
The Weitzenb\"{o}ck connection $T_{\lambda\mu\nu}$ can be defined by the following three
parts \cite{HS79}
\begin{equation}
T_{\lambda\mu\nu}= \frac{1}{2}\big( t_{\lambda\mu\nu}
-t_{\lambda\nu\mu} \big)+
 \frac{1}{3}\big(g_{\lambda\mu}V_{\nu} - g_{\lambda\nu}V_{\mu} \big) + \varepsilon_{\lambda\mu\nu\rho}A^{\rho},
\end{equation}
where
$$
t_{\lambda\mu\nu}=\frac{1}{2}\big(T_{\lambda\mu\nu}+T_{\mu\lambda\nu}\big)
+\frac{1}{6}\big(g_{\nu\lambda}V_{\mu}+g_{\mu\nu}V_{\lambda}\big)
-\frac{1}{3}g_{\lambda\mu}V_{\nu}
$$
is the tensor part that represents the torsion tensor,
\begin{equation}\label{V}
V_{\mu}=T^{\nu}_{\,\,\nu\mu}
\end{equation}
is the vector part that gives the torsion vector, and
\begin{equation} \label{ax}
A^{\mu}=h_{a}^{\,\,\mu}A^{a}=\frac{1}{6}\varepsilon^{\mu\nu\rho\sigma}T_{\nu\rho\sigma}
\end{equation}
is the axial-vector part that  defines  the torsion axial-vector,
which represents the axial symmetry deviation from spherical symmetry \cite{NH80}.\\

$\varepsilon^{\mu\nu\rho\sigma}$ and
$\varepsilon_{\mu\nu\rho\sigma}$ are  completely antisymmetric tensors with respect to the coordinates
basis and defined by \cite{M52}
$$
\varepsilon^{\mu\nu\rho\sigma}=\frac{1}{\sqrt{-g}}\delta^{\mu\nu\rho\sigma},
$$
$$
\varepsilon_{\mu\nu\rho\sigma}=\sqrt{-g}\delta_{\mu\nu\rho\sigma},
$$
where $\delta^{\mu\nu\rho\sigma}$ and $\delta_{\mu\nu\rho\sigma}$
are the  antisymmetric tensor densities of weight $-1$ and
$+1$, respectively, with normalization $\delta_{0123}=-1$ and $\delta^{0123}=+1$.\\
In the presence of matter, the action of teleparallel gravity  is
given by

$$
{\A }= {\frac{1}{16\pi}} \int d{^4}x h S^{\rho\mu\nu}T_{\rho\mu\nu}+\int
d^4x h \pounds_{M},
$$
where $h=det(h^a_{\,\,\mu})$, $\pounds_{M}$ is the Lagrangian of a
source field and
\begin{equation}\label{S}
S^{\rho\mu\nu}= c_{1}T^{\rho\mu\nu} +
\frac{c_{2}}{2}\big(T^{\mu\rho\nu} - T^{\nu\rho\mu}\big) +
\frac{c_{3}}{2}\big(g^{\rho\nu}T^{\sigma\mu}_{\,\,\,\,\sigma} -
g^{\mu\rho}T^{\sigma\nu}_{\,\,\,\,\sigma}\big),
\end{equation}
is a tensor written in terms of the torsion of the
Weitzenb\"{o}ck connection. In the above form $c_{1}, c_{2}$ and
$c_{3}$ are the three dimensionless coupling constants of Teleparallel Gravity.\\
For the so called teleparallel equivalent of general relativity,
the specific choice for these constants is determined by \cite{HS79}
\begin{equation}\label{2.13}
c_{1} =\frac{1}{4}, \qquad c_{2}=\frac{1}{2}, \qquad c_{3}=-1.
\end{equation}
The energy-momentum complexes of Einstein, Bergmann-Thomson  and
Landau-Lifshitz in Teleparallel Gravity, respectively, are given
by \cite{{M62}}
\begin{equation}\label{EBL}
\begin{array}{ccc}
hE^{\mu}_{\,\,\,\,\nu} & =
\frac{1}{4\pi}\partial_{\lambda}\Big(\mho_{\nu}^{\,\,\mu\lambda}
\Big),\\
hB^{\mu\nu} & = \frac{1}{4\pi}\partial_{\lambda}\Big(g^{\mu\beta}\mho_{\beta}^{\,\,\nu\lambda}\Big),\\
hL^{\mu\nu} & =
\frac{1}{4\pi}\partial_{\lambda}\Big(hg^{\mu\beta}\mho_{\beta}^{\,\,\nu\lambda}\Big),
\end{array}
\end{equation}
where $\mho_{\nu}^{\,\,\mu\lambda}$ is the Freud's super-potential
and defined as follows
\begin{equation}\label{U}
\mho_{\nu}^{\,\,\mu\lambda}=hS_{\nu}^{\,\,\mu\lambda}.
\end{equation}
 The energy and momentum distributions in
the above complexes, respectively, are
\begin{equation}\label{e-mD}
\begin{array}{ccc}
^{E}P_\mu & = \int_{\Sigma}hE^0_{\,\,\mu}d^3x,\\
^{BT}P_\mu & = \int_{\Sigma}hB^0_{\,\,\mu}d^3x,\\
^{LL}P_\mu & = \int_{\Sigma}hL^0_{\,\,\mu}d^3x,
\end{array}
\end{equation}
where $P_{0}$ is the energy, $P_{i}\quad (i=1,2,3)$ are the momentum
components and the integration hypersurface $\Sigma$ is described by
$x^0 =t$ constant.

\subsection{Energy-momentum density in teleparallel gravity}
In this section, energy-momentum density is calculated for the space-time (\ref{VS}) in the context of the theory of
teleparallel gravity. \\
Tetrad construction is a simple procedure, if the given metric tensor, $g_{\mu\nu}$, is in diagonal form. One should only make sure that this tetrad satisfies the  conditions (\ref{g}), (\ref{h}) and  (\ref{h-g}). If the space-time  has a non-diagonal form,
 there is some complexity in the construction of a tetrad, which corresponds to $g_{\mu\nu}$.\\
Our construction of the orthogonal  tetrad  to the metric (\ref{VS}) is in the following form (see Gad \cite{Arxiv2020})
\begin{equation}\label{FT}
{h^a_{\,\,\mu}} =
\left[ \begin{array}{cccc}
\sqrt{\frac{f\ell +m^2}{\ell}} & 0 & 0 & 0 \\
0 & e^{\psi} & 0 & 0 \\
0 & 0 & e^{\psi} & 0 \\
-\frac{m}{\sqrt{\ell}} & 0 & 0 & \sqrt{\ell}
\end{array} \right],
\end{equation}
where $h:=\det h^a_{\,\,\mu}= \sqrt{f\ell +m^2}e^{2\psi}=\sqrt{-g}$
and its inverse, ${h_a^{\,\,\mu}}= g^{\mu\nu}\eta_{ab}{h^b_{\,\,\nu}}$ is
\begin{equation}\label{IFT}
{h_a^{\,\,\mu}} =
\left[ \begin{array}{cccc}
\sqrt{\frac{\ell}{f\ell + m^2}} & 0 & 0 & 0 \\
0 & e^{-\psi} & 0 & 0 \\
0 & 0 & e^{-\psi} & 0 \\
\frac{m}{\sqrt{\ell(f\ell +m^2)}} & 0 & 0 & \frac{1}{\sqrt{\ell}}
\end{array} \right].
\end{equation}
Using the above components of $h^a_{\,\,\mu}$ and $h_a^{\,\,\mu}$ in (\ref{conn}),
we find the Weitzenb\"{o}ck connection, whose  nonvanishing components are (see Gad \cite{Arxiv2020})
\begin{equation}\label{Fc}
\begin{array}{ccc}
  \Gamma^0_{\,\,00} &= &\frac{\ell}{2(f\ell +m^2)}\Big( \frac{\dot{f}\ell+f\dot{\ell}+2m\dot{m}}{\ell} - \frac{(f\ell +m^2)\dot{\ell}}{\ell^2}\Big),\\
\Gamma^0_{\,\,01} &= &\frac{\ell}{2(f\ell +m^2)}\Big( \frac{f^{\prime}\ell+f\ell^{\prime}+2m m^{\prime}}{\ell} - \frac{(f\ell +m^2)\ell^{\prime}}{\ell^2}\Big),\\
\Gamma^1_{\,\,10}& = &\Gamma^2_{\,\,20} = \dot{\psi} , \quad \Gamma^1_{\,\,11}=\Gamma^2_{\,\,21} =  \psi^{\prime}, \quad \Gamma^3_{\,\,30} = \frac{\dot{\ell}}{2\ell}, \quad \Gamma^3_{\,\,31}= \frac{\ell^\prime}{2\ell},\\
\Gamma^3_{\,\,00} &=& \frac{m}{2(f\ell +m^2)}\Big( \frac{\dot{f}\ell+f\dot{\ell}+2m\dot{m}}{\ell} - \frac{(f\ell +m^2)\dot{\ell}}{\ell^2}\Big)- \frac{\dot{m}}{\ell}+\frac{m\dot{\ell}}{2\ell^2},\\
\Gamma^3_{\,\,01} &=& \frac{m}{2(f\ell +m^2)}\Big( \frac{f^{\prime}\ell+f\ell^{\prime}+2m m^{\prime}}{\ell} - \frac{(f\ell +m^2)\ell^{\prime}}{\ell^2}\Big)-\frac{m^\prime}{\ell}+\frac{m\ell^\prime}{2\ell^2} ,
\end{array}
\end{equation}
where prime denotes  a partial derivative with respect to $x$ and dot denotes
a partial derivative with respect to $t$.\\
Inserting (\ref{Fc}) into (\ref{tor}) yields the following nonvanishing torsion components (see Gad \cite{Arxiv2020})
\begin{equation}\label{T}
\begin{array}{ccc}
T^0_{\,\,01}= -T^0_{\,\,10}= -\Gamma^0_{\,\,01}, \quad T^1_{\,\,10}= -T^1_{\,\,01}= -\Gamma^1_{\,\,10} \\ T^2_{\,\,20}= -T^2_{\,\,02}= -\Gamma^2_{\,\,20}, \quad
T^2_{\,\,21}= -T^2_{\,\,12}= -\Gamma^2_{\,\,21}, \\ T^3_{\,\,01}= -T^3_{\,\,10}= -\Gamma^3_{\,\,01}, \quad T^3_{\,\,30}= -T^3_{\,\,03}= -\Gamma^3_{\,\,30},\\
T^3_{\,\,31}= -T^3_{\,\,13}= -\Gamma^3_{\,\,31}.
\end{array}
\end{equation}
Inserting (\ref{T}) into (\ref{S}), using (\ref{2.13}),
the non-vanishing required components of $S_\beta^{\,\,\mu \nu}$ are
\begin{equation}
\begin{array}{cccc}
 S_0^{\,\,01} &= & -\frac{1}{4 ({f\ell+m^2})}\Big(mm^\prime+f\ell^\prime +2(\ell f+m^2)\psi^\prime\Big),\\
 S_1^{\,\,01}& = & -\frac{2\ell\dot{\psi}+\dot{\ell}}{4 ({f\ell+m^2})},\\
  S_3^{\,\,01}&=  &\frac{(\ell m^\prime-\ell^\prime)e^{-2\psi}}{4 ({f\ell+m^2})},\\
  S_0^{\,\,30} &=  &-\frac{m\dot{\psi}}{ {f\ell+m^2}},\\
 S_3^{\,\,30} &=  &\frac{\ell\dot{\psi}}{ {f\ell+m^2}},\\
 S_0^{\,\,31}& = & \frac{( mf^\prime-fm^\prime)e^{-2\psi}}{4 ({f\ell+m^2})},\\
  S_3^{\,\,31}& = &\frac{e^{-2\psi}}{4 ({f\ell+m^2})}(\ell f^\prime + mm^\prime +2\psi^\prime(\ell f +m^2)).
\end{array}
\end{equation}
Inserting  these components  into equations (\ref{U}), using
(\ref{EBL}),   we get the energy-momentum density in the
prescriptions of Einstein,
Bergmann-Thomson and Landau-Lifshitz,
respectively,
\begin{equation}
\begin{array}{lcl}
hE^{0}_{0}=-\frac{1}{4\kappa ({f\ell+m^2})^{\frac{3}{2}}}\Big((f\ell+m^2)(4(f\ell+m^2)\psi^{\prime\prime}+2mm^{\prime\prime}+4mm^\prime\psi^\prime+f^\prime\ell^\prime+2f\ell^{\prime\prime}\\
+2\ell f^\prime\psi^\prime+2f\ell^\prime\psi^\prime)-f^2\ell^{\prime 2}+m^2 f^\prime\ell^\prime+2f\ell m^{\prime 2}-3fmm^\prime\ell^\prime-\ell f^\prime mm^\prime\Big),\\
hE^{0}_{1} =-\frac{e^{2\psi}}{4\kappa ({f\ell+m^2})^{\frac{3}{2}}}\Big(2(f\ell+m^2)(4\ell\dot{\psi}\psi^\prime+\ell^\prime\dot{\psi}+2\dot{\ell}\psi^\prime+2\ell\dot{\psi}^\prime+\dot{\ell}^\prime)+2m^2\ell^\prime\dot{\psi}-\\
 2\ell^2\dot{\psi}f^\prime-4\ell mm^\prime\dot{\psi}-f\dot{\ell}\ell^\prime-\ell\dot{\ell}f^\prime-2\dot{\ell}mm^\prime\Big),\\
hE^{0}_{3} = \frac{1}{4\kappa ({f\ell+m^2})^{\frac{3}{2}}}\Big(2(f\ell+m^2)(\ell m^{\prime\prime}-m\ell^{\prime\prime})+fm\ell^{\prime 2}+m\ell\ell^\prime f^\prime+2m^2 m^\prime\ell^\prime\\
-f\ell\ell^\prime m^\prime -\ell^2 f^\prime m^\prime -2\ell mm^{\prime 2}\Big).
\end{array}
\end{equation}

\begin{equation}
\begin{array}{lcl}
hB^{00}=\frac{1}{4\kappa ({f\ell+m^2})^{\frac{3}{2}}}\Big(-2(f\ell+m^2)(\ell^\prime\psi^\prime +2\ell\psi^{\prime\prime}+\ell^{\prime\prime})-2m^2\ell^\prime\psi^\prime+2\ell^2 f^\prime\psi^\prime+4\ell mm^\prime\psi^\prime\\
 +f\ell^{\prime 2}+f^\prime\ell\ell^\prime+2\ell^\prime mm^\prime\Big),\\
hB^{01}=\frac{1}{4\kappa ({f\ell+m^2})^{\frac{3}{2}}}\Big(2(f\ell+m^2)(\ell^\prime\dot{\psi} +2\ell\dot{\psi}^\prime+\dot{\ell}^\prime)+2m^2\ell^\prime\dot{\psi}-2\ell^2 f^\prime\dot{\psi}-4\ell mm^\prime\dot{\psi}\\
-f\ell^\prime\dot{\ell}-f^\prime\ell\dot{\ell}-2mm^\prime\dot{\ell}\Big),\\
 hB^{02} = 0,\\
 hB^{03}= \frac{1}{4\kappa ({f\ell+m^2})^{\frac{3}{2}}}\Big(-2(f\ell+m^2)(2m\psi^{\prime\prime}+m^{\prime\prime}) +2fm\ell^\prime\psi^\prime+2m\ell f^\prime\psi^\prime-4f\ell m^\prime\psi^\prime\\
+f\ell^\prime m^\prime+\ell f^\prime m^\prime+2mm^{\prime 2}\Big).
\end{array}
\end{equation}
\begin{equation}
\begin{array}{lcl}
hL^{00}=-\frac{e^{2\psi}}{2\kappa}\Big(4\ell\psi^{\prime 2}+2\ell\psi^{\prime\prime}+4\ell^\prime\psi^\prime+\ell^{\prime\prime}\Big),\\
hL^{01} =\frac{e^{2\psi}}{2\kappa}\Big(4\ell\dot{\psi}\psi^\prime+2\ell\dot{\psi}^\prime+ 2\ell^\prime\dot{\psi}+2\dot{\ell}\psi^\prime+\dot{\ell}^\prime\Big),\\
hL^{02} = 0,\\
hL^{03}=-\frac{e^{2\psi}}{2\kappa}\Big(4m\psi^{\prime 2} +2m\psi^{\prime\prime}+4m^\prime\psi^\prime+m^{\prime\prime}\Big).
\end{array}
\end{equation}
These  are consistent with the results obtained
 in the context of General Relativity (see section 2).

\section{Conclusion}

This paper is focused, firstly, on the study of the energy-momentum density for the gravitational field
of Van Stockum  space-time in GR by applying  the energy-momentum complexes of Einstein, M{\o}ller, Bergmann-Thomson and Landau-Lifshitz, using double index complexes. The expressions for the energy-momentum density are well-defined in all the aforesaid prescriptions and do not give the same results. The momentum component in the $y$-direction is zero in all four prescriptions used. The second aim of this paper was to study the same issue in TPG using the Einstein, Bergmann-Thomson and Landau–Lifshitz  complexes. As in the case of GR these prescriptions do not give the same results for the energy-momentum density, while every complex gave the same corresponding results  obtained in GR\footnote{Finding energy-momentum density using M{\o}ller prescription in the theory of teleparallel gravity will be postponed to another article.}. This  sustains that the theories of TPG and GR are  equivalent. Moreover, we have shown that for all homogeneous space-times, with metric components $g_{\mu\nu}$ are functions of time variable, $t$, alone and independent of space variables, the total gravitational energy for any finite volume is identically zero. To sustain this result, we assume that the scale factors of the space-time (\ref{VS}) are functions in $t$ only. We find that the energy-momentum density is identically zero. This sustains theorem (\ref{th}).


\end{document}